\documentclass{ifacconf}

\usepackage{graphicx}      
\usepackage{natbib}        
\usepackage{xurl}
\usepackage{svg}
\begin{document}
\begin{frontmatter}

\title{Bringing AI Assistance to the Beznau Nuclear Power Plant: A Hierarchical System of LLM Agents for Reactor Monitoring} 

\thanks[footnoteinfo]{We thank Axpo Power AG for their support of this collaboration.}

\author[First]{Wenqi Zhou} 
\author[First]{Leonardo Baldo} 
\author[Third]{Edwin Kolbe}
\author[First,Fourth]{Manuel Arias Chao}

\address[First]{Institute for Data Science, Zurich University of Applied Science, Winterthur, 8401, Switzerland.}
\address[Third]{Department of Physics, University of Basel, Basel, 4056, Switzerland.}
\address[Fourth]{Faculty of Aerospace Engineering, Delft University of Technology, HS 2926 Delft, The Netherlands.}

\begin{abstract}                
Beznau Unit 1 in Switzerland, in operation since 1969, is the most experienced nuclear reactor still running worldwide. Five decades of incremental modernization have left its monitoring systems, alarm logs, variable catalogs, technical specifications and investigation reports in separate silos that were never designed to be queried together. As a result, investigating a core instrumentation anomaly means slow, manual gathering of knowledge scattered across databases, documents, and experienced colleagues. We present a hierarchical multi-agent assistant built on large language models (LLMs) to support these investigations and reduce the need for tedious manual searches in SAP systems. An orchestrating agent breaks down an engineer's question and hands the parts to specialized sub-agents, each with read-only access to one source. Domain expertise is written as natural-language procedures that plant engineers can read, review and revise. The system is strictly advisory, with every answer traced back to its sources. We also report organizational lessons from introducing agentic AI into a culture built on safety, traceability and conservative change. Beyond the technical architecture, the Beznau case study examines how an assistant can be introduced into a setting where evidence must be traceable, changes carefully reviewed, and operational decisions retained by plant personnel, paving the way for the responsible use of agentic AI in industrial plant monitoring.
\end{abstract}

\end{frontmatter}

\section{Introduction}

Investigating an instrumentation anomaly in a complex industrial system requires engineers to establish what a measurement represents and how its behaviour relates to the operating conditions of the plant. Numerical readings provide the starting point but diagnosis requires a much broader body of knowledge. Earlier tests and measurements, maintenance records, previous investigation reports, and technical specifications are just some of the pivotal information sources that guide the reasoning process (\cite{kang2026maintenance}). Much of the work then consists of gathering this evidence across several applications and determining how it relates to the anomaly under investigation.
In legacy plants that have operated for several decades, this task involves information accumulated through successive equipment modifications and changes in monitoring systems. The resulting layered architecture reflects decades of incremental modernisation that have supported continued operation while making troubleshooting increasingly demanding. Personnel know-how in dealing with these issues is hence an essential part of the plant's engineering knowledge.
Recent energy crises and concerns over security of supply have renewed interest in extending the operating lives of existing nuclear power plants, with several countries reconsidering planned closures (\cite{iea2022Nuclear}). Longer operation increases the importance of transferring accumulated experience to the personnel who will maintain and operate these plants in the coming decades. It is no coincidence that the International Atomic Energy Agency (IAEA) has identified the preservation and transfer of such knowledge as a continuing concern for nuclear operating organisations (\cite{iaea2006Knowledge}).

The growing use of large language models in engineering is changing how personnel interact with technical information. 
An engineer can express a question in the language used to discuss a problem with a colleague, while an assistant retrieves the records and performs the operations required to address it. This capability is particularly relevant to diagnostic work, where a question often crosses the boundaries between monitoring data, maintenance history, and technical documentation. 
Retrieval-augmented generation (RAG) provides a technical basis for this interaction by supplying an LLM with material retrieved from external sources. Responses can therefore incorporate information maintained independently of model training (\cite{lewis2020RAG}). Tools using agents extend this approach by selecting operations and using the returned results to guide subsequent steps (\cite{yao2023ReAct}). The investigation can hence develop as evidence is obtained, with each result informing the next task. 

Industrial maintenance research has begun to examine these capabilities in settings where operational data and engineering records are maintained in different forms. For instance, MindRAG combines multimodal retrieval with LLM-based agents to support condition-monitoring workflows using measurements, annotations, and maintenance work orders (\cite{lowenmark2025MindRAG}). 
MAKG introduces collaborative agents and knowledge-graph retrieval for equipment-maintenance questions (\cite{lin2026MAKG}).  These studies show how the organisation of evidence and the design of the agent workflow influence the assistance provided.
The idea of recreating a team of specialists through agents has found fertile ground in many applications, and strategies based on this paradigm are indeed flourishing. In this sense, orchestrated multi-agent architectures organise an investigation around specialised agents whose work is directed by a coordinator. The coordinator interprets the request, assigns tasks, and uses the returned results to guide further work. Magentic-One provides a general implementation of this approach, with an orchestrator responsible for planning, tracking progress, and recovering from unsuccessful steps (\cite{fourney2024MagenticOne}). In industrial maintenance, Farahani et al. use an LLM planner to coordinate specialised agents across data preparation, analysis, and maintenance recommendation (\cite{farahani2026Hybrid}). Gong et al. examine how diagnostic questions can be routed according to their complexity, involving several agents in deliberation when further reasoning is required (\cite{gong2026SensorFailure}). Schall addresses the integration of monitoring agents into process automation through an orchestrator and a separate validation layer grounded in plant documentation (\cite{schall2026SafeIntegration}). 

In nuclear engineering, RADIANT-LLM represents one of the most recent and advanced framework, addressing retrieval from technical documentation through a multimodal architecture that preserves page and figure source (\cite{ndum2027RADIANT}). Reactor-operation assistance has also been investigated in simulation environments. Lee et al. examine an LLM agent performing tasks within a reactor simulator (\cite{lee2025ReactorAgent}), while AROMA-GPT connects an assistant to a physics-based digital twin through dedicated analytical tools (\cite{ndum2026AROMA}). Lim et al. extend this line of work to a physical thermal-fluid experimental facility equipped with a digital twin and an LLM interface (\cite{lim2025ThermalFluid}).
These studies demonstrate increasingly integrated forms of LLM support for nuclear engineering, although their validation remains within document-based, simulated, or experimental environments. The present work moves this line of research into an operational plant setting, where the assistant is deployed and used within the investigation workflow of an existing commercial nuclear power plant.


Applying these capabilities in such an environment introduces a different set of requirements. Plant investigations depend on the relationship between heterogeneous records, operational data, maintenance information, technical documentation, and the accumulated knowledge of plant personnel. At the same time, deployment in a safety-critical and highly regulated environment requires clear source traceability, controlled access to plant information, reproducible reasoning steps, and preservation of human authority over the final engineering judgement. These requirements are amplified in a legacy plant, where decades of equipment history and operational experience coexist.
This combination motivates the coordinated and orchestrated workflow considered in this paper. 

\begin{figure}
    \centering
    \includegraphics[width=8.4cm]{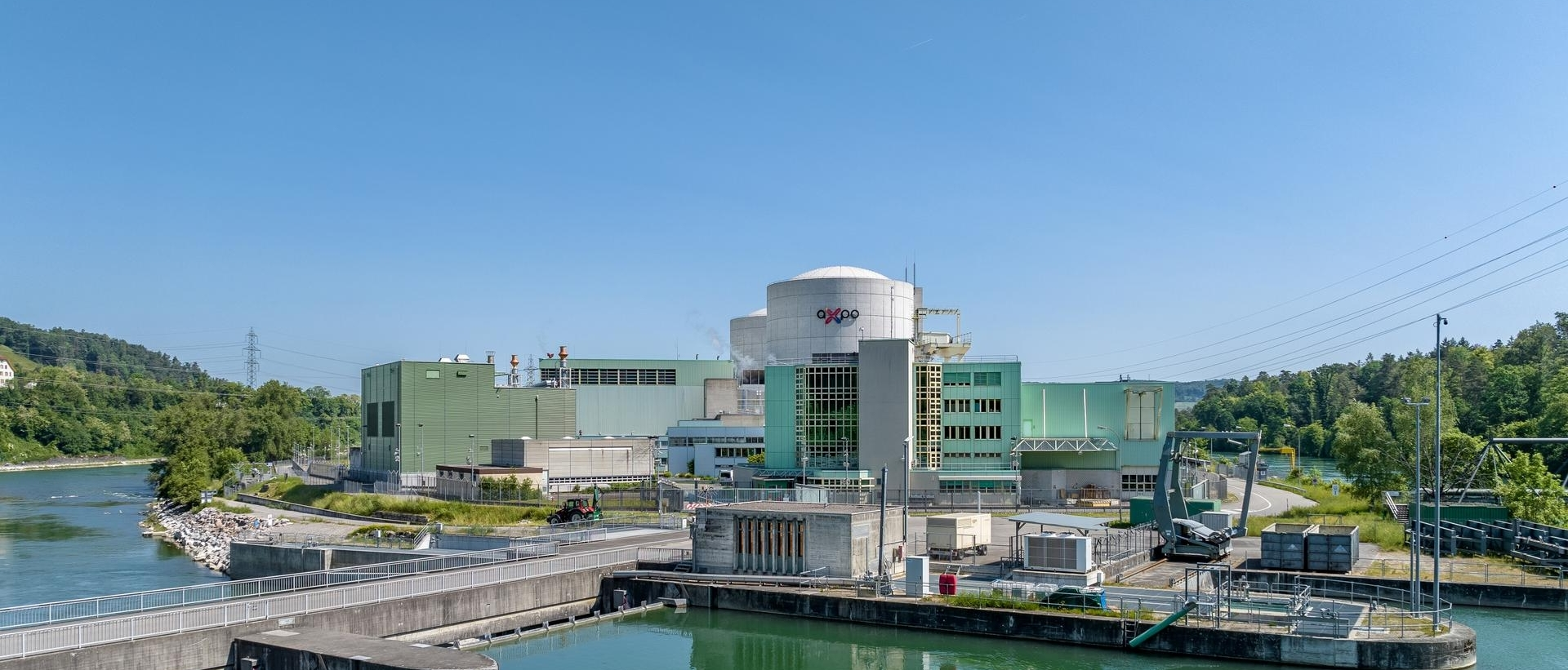}
    \caption{Beznau Nuclear Power Plant (NPP). (Copyright Dani Werder for Axpo).}
    \label{fig:Beznau}
\end{figure}

The Beznau Nuclear Power Plant (Figure \ref{fig:Beznau}) in Switzerland provides a particularly demanding operational test bed for this problem. Unit 1 has been in operation since 1969 and is recognised as the oldest nuclear power reactor still in operation worldwide (\cite{axpo2019Beznau}). Its operating history spans several generations of equipment and engineering personnel, together with substantial investment in modernisation. A plant information and monitoring system was introduced in 1991 and modernised in 2015. It processes thousands of signals per unit to support monitoring and recording of operating history. A digital reactor protection system was introduced in 2001 (\cite{axpo2019Beznau}). These upgrades have added modern monitoring capabilities to a plant with decades of accumulated operational knowledge. Beznau offers a concrete setting in which to examine how an assistant can bring this information into an investigation while retaining the context needed to interpret it.

This paper presents a hierarchical multi-agent assistant for reactor-monitoring investigations at Beznau. An orchestrating agent coordinates specialised agents with read-only access to plant sources and assembles their findings into an advisory response linked to supporting evidence. Plant engineers review the natural-language procedures governing the agents' work and retain operational judgement.

The main contributions are:
\begin{itemize}
    \item An orchestrated architecture that brings operational data and engineering records from several plant sources into a common investigation workflow, implemented in a pilot environment.
    \item The use of natural-language domain procedures that plant engineers can read, review, and revise.
    \item A monitoring case study illustrating the coordination of agent tasks and the presentation of evidence for engineering assessment.
    \item Lessons from introducing and deploying agentic assistance into established nuclear engineering practice, with attention to traceability and the review of generated interpretations.
\end{itemize}

\section{Sources and Concept of Operations}
Reactor-monitoring investigations at Beznau rely on operational data and engineering records maintained in several information systems:

\begin{itemize}
    \item \textbf{Operational data:} ANIS (\emph{Anlageinformationssystem}, or plant information system) collects and records plant signals for monitoring and historical analysis (\cite{axpo2019Beznau}). The GARDEL system provides online reactor core monitoring (\cite{studsvikGARDEL}). 

    \item \textbf{Alarm logs.}

    \item \textbf{Variable definitions.}

    \item \textbf{Technical specifications.}

    \item \textbf{Previous investigation reports:} These document earlier events, the evidence examined, and the engineering assessments reached, providing a reference for comparable observations.

    \item \textbf{Organisation database:} This records responsibilities within the plant organisation.
\end{itemize}




The assistant supports an engineer investigating a monitoring observation through a conversational interface. In fact, a question leads to the retrieval of plant information and the analysis of measurements. The investigation develops according to the evidence obtained, with the engineer able to refine the question or request further analysis.
The question is directly handled by the orchestrator which coordinates retrival and analysis. The different resources  are retrieved in read-only mode and analysed to identify relevant passages and to perform specific analysis. 
After the analysis the response is sent back to the engineer through the conversational interface and can inspect its supporting evidence. A follow-up question may refine the question or request clarification of an interpretation. This interaction allows the investigation to progress as the engineer assesses the findings. The resulting response is advisory and remains subject to assessment by plant personnel.



\section{Orchestrated Multi-Agent Architecture}
\label{sec:architecture}

\begin{figure}
    \centering
    \includegraphics[width=8.4cm]{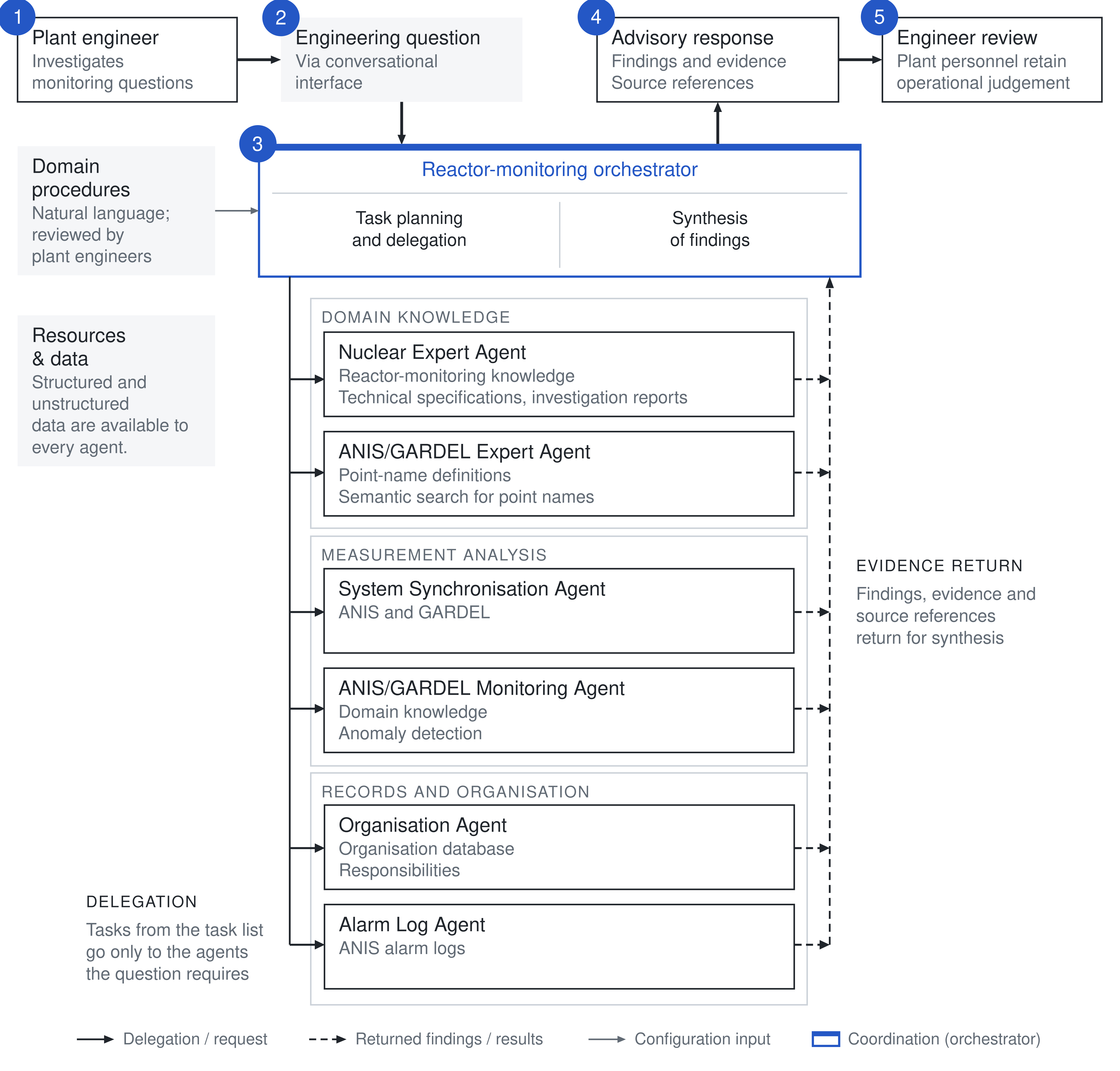}
    \caption{Orchestration of the reactor-monitoring assistant. The orchestrator plans the investigation and delegates tasks to the specialist agents required by the engineer's question.}
    \label{fig:orchestration}
\end{figure}

The framework is organised around an orchestrating agent and six specialist agents, as shown in Fig.~\ref{fig:orchestration}. The engineer submits a question through the conversational interface. The orchestrator interprets the request, develops a task list, and delegates the work to the agents whose expertise and information sources are relevant to the investigation. The orchestrator brings their findings together into an advisory response supported by source references.

\begin{figure*}[t]
    \centering
    \includegraphics[width=0.65\textwidth]{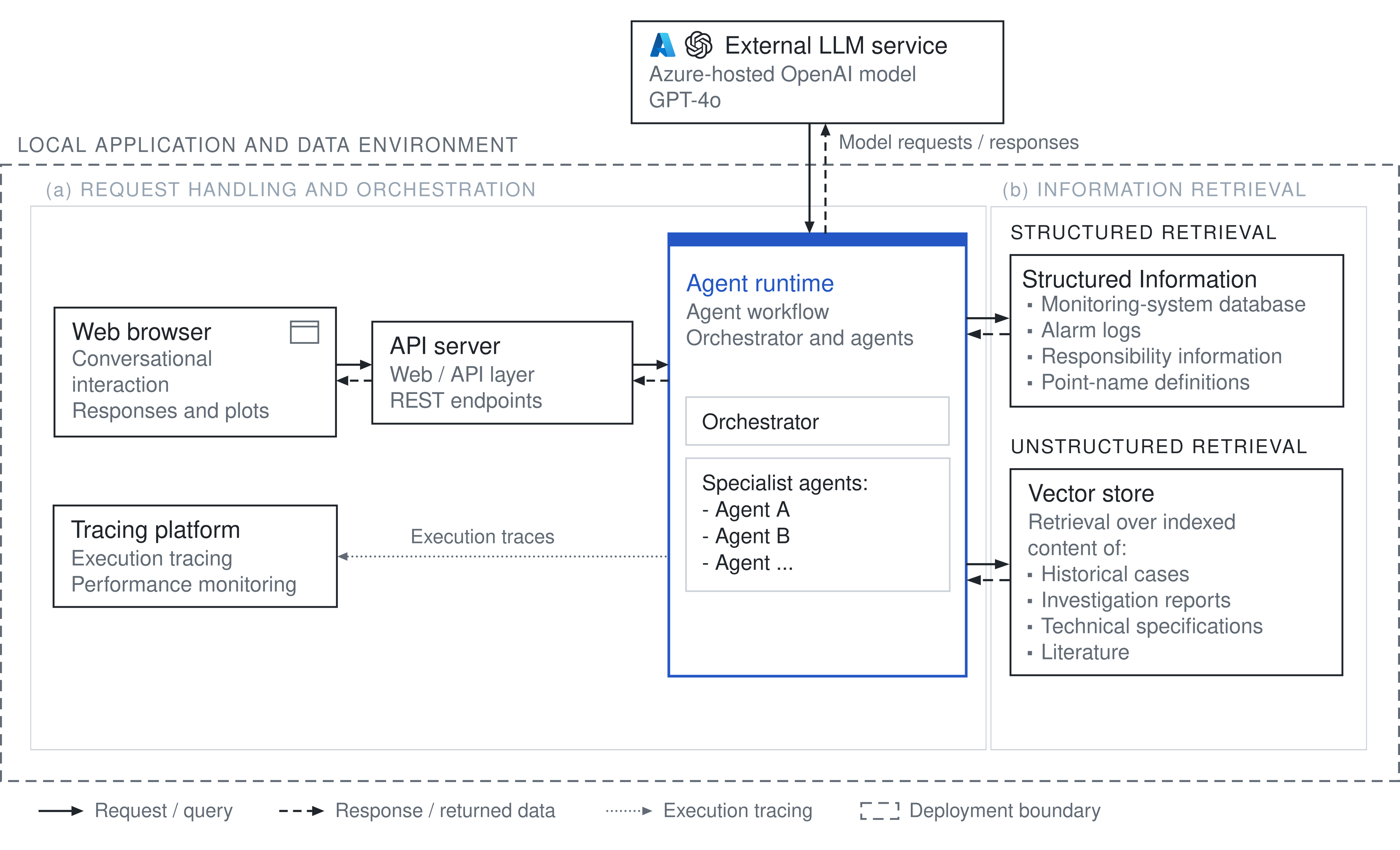}
    \caption{Software architecture of the reactor-monitoring assistant application.}
    \label{fig:software_architecture}
\end{figure*}

\textbf{Investigation Planning and Coordination.}
An engineering question often contains several dependent tasks. Investigating an unexpected signal fluctuation, for example, may first require identifying the corresponding point name and retrieving its measurement history. The observed behaviour can then be examined alongside associated alarms and previous investigation reports. The orchestrator coordinates these steps, using the information returned by one agent to establish the context for subsequent requests.
Delegation follows the needs of the investigation and domain procedures defined by the plant experts. A request concerning responsibility information may involve only the \textbf{Reactor-Monitoring Orchestrator (RMO)}, whereas an instrumentation anomaly may require contributions from several specialists. 
The specialist agents have the following roles:

\begin{itemize}
    \item \textbf{Nuclear Expert Agent (NEA):} Consults reactor-monitoring knowledge, technical specifications, and investigation reports to provide engineering context for the observed behaviour.

    \item \textbf{Alarm Log Agent (ALA):} Retrieves relevant entries from the ANIS alarm logs to establish the events surrounding the observation.

    \item \textbf{Organisation Agent (OA):} Queries the database containing the organizational charts to identify responsibilities and the personnel associated with a technical issue.

    \item \textbf{System Synchronisation Agent (SSA):} Examines synchronisation between ANIS and GARDEL data to support comparison of their measurements.

    \item \textbf{ANIS/GARDEL Expert Agents (AEA/GEA):} Retrieve point-name definitions through semantic search, helping identify the variables required for the investigation.

    \item \textbf{ANIS/GARDEL Monitoring Agents (AMA /GMA):} Appy monitoring procedures and analytical tools to examine measurement behaviour and detect anomalies.
\end{itemize}

This division gives each agent a defined scope and access to the resources needed for its task. Coordination remains with the orchestrator, which assembles the individual contributions in relation to the original engineering question.

\textbf{Domain Procedures.} Domain procedures guide the investigation and the work assigned to the specialist agents. They are expressed in natural language and describe how plant information should be retrieved and examined for a given monitoring task. Plant engineers can read, review, and revise these instructions as operational experience develops or information systems change. They guide the selection of relevant variables, the consultation of supporting records, and the interpretation of analytical results.

\textbf{Evidence Assembly and Engineering Review.}
Specialist agents return findings together with the evidence and source references supporting them. The orchestrator assembles these contributions into a response that addresses the engineer's question. Measurement histories and analytical results establish the observed behaviour, while retrieved records provide the context for its interpretation.
The response presents the findings and supporting references for engineering review. Where the investigation suggests a follow-up action, the assistant can identify relevant documentation or the personnel responsible for the issue. Plant personnel assess the interpretation and determine the operational response. The architecture thus supports the investigation from the initial question to the examination of evidence, with operational judgement retained by the engineer.

\textbf{Software Architecture.}
Figure~\ref{fig:software_architecture} shows the software implementation of the assistant. A browser interface exchanges requests, responses, and plots with the assistant backend through a REST API. Agent runtime manages the orchestrator and specialist-agent workflows, including task delegation and the return of results.
The backend accesses structured plant information through database queries and retrieves indexed documents from a vector store. Language-model requests are sent to GPT-4o through an Azure OpenAI service. Using an Azure-hosted deployment in Switzerland rather than the public OpenAI API keeps prompts and responses within a controlled enterprise environment. This supports the confidentiality requirements associated with sensitive plant data. The application and its data stores reside within the local deployment environment, while a locally deployed tracing platform records execution traces and performance information to support inspection of the agent workflow.

\section{Monitoring Scenarios and Evaluation}

The following scenarios illustrate how the Reactor Monitoring Assistant combines specialised agents to address questions requiring operational data and plant documentation. The examples cover an introduction to the assistant, a query about anomaly detection, and an investigation of fluctuations in ANIS data. For demonstration purposes, we ingested synthetic fault, which is shown in the second example. The fluctuation examined in the third example is a known phenomenon that has been documented and is recognized by the manufacturer, which does not pose a risk to plant operation and was selected solely to demonstrate the assistant's capability. Each response identifies the agents engaged and provides a \textit{View references} option through which the supporting sources can be retrieved. These features allow the engineer to inspect the evidence underlying the answer and assess its interpretation.

\begin{figure}
    \centering
    \includegraphics[width=8.2cm]{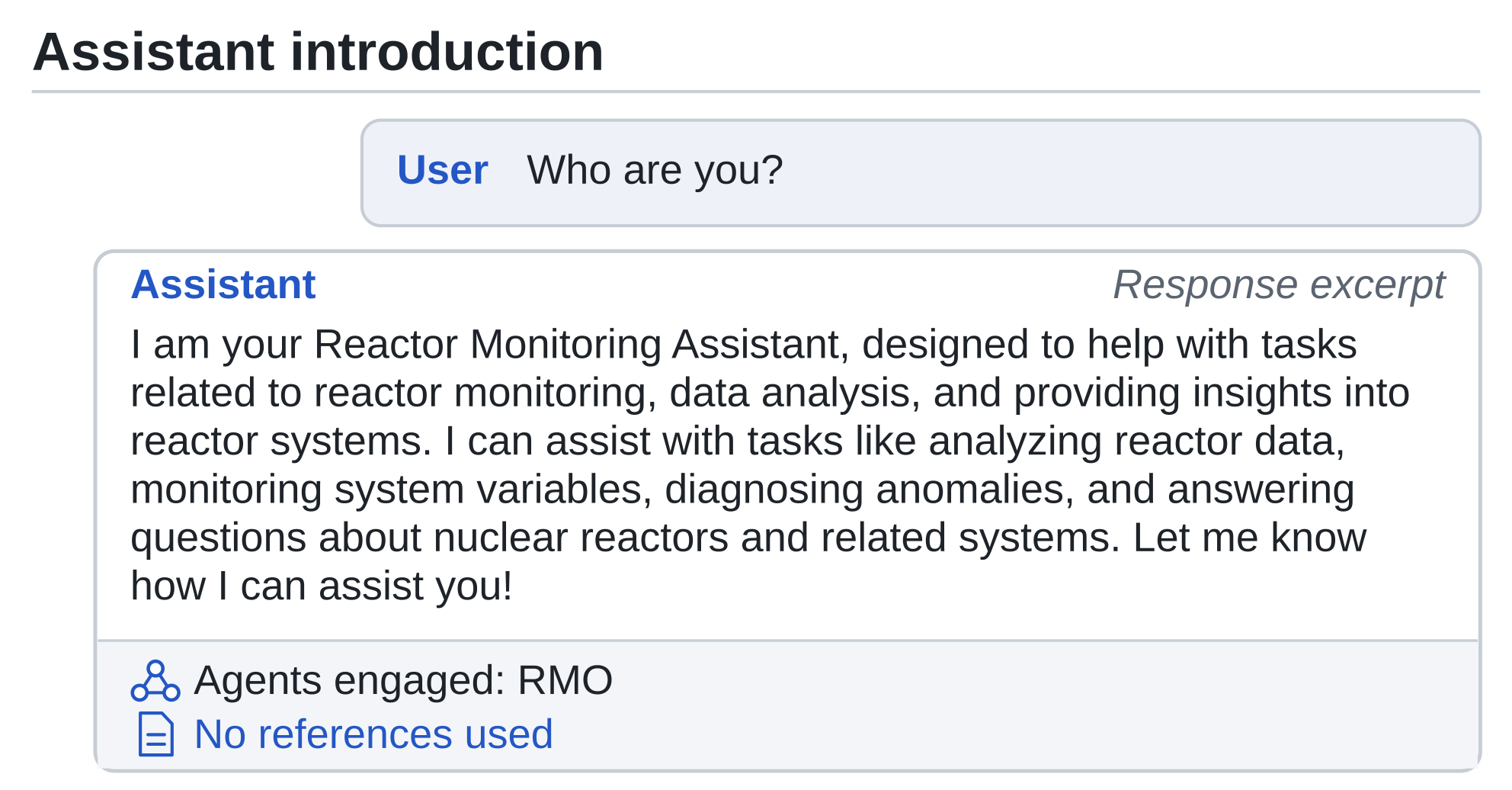}
    \caption{Introduction of the Reactor Monitoring Assistant, showing its stated scope, the agent engaged, and access to supporting references.}
    \label{fig:assistant_introduction}
\end{figure}

\begin{figure}
    \centering
    \includegraphics[width=8.4cm]{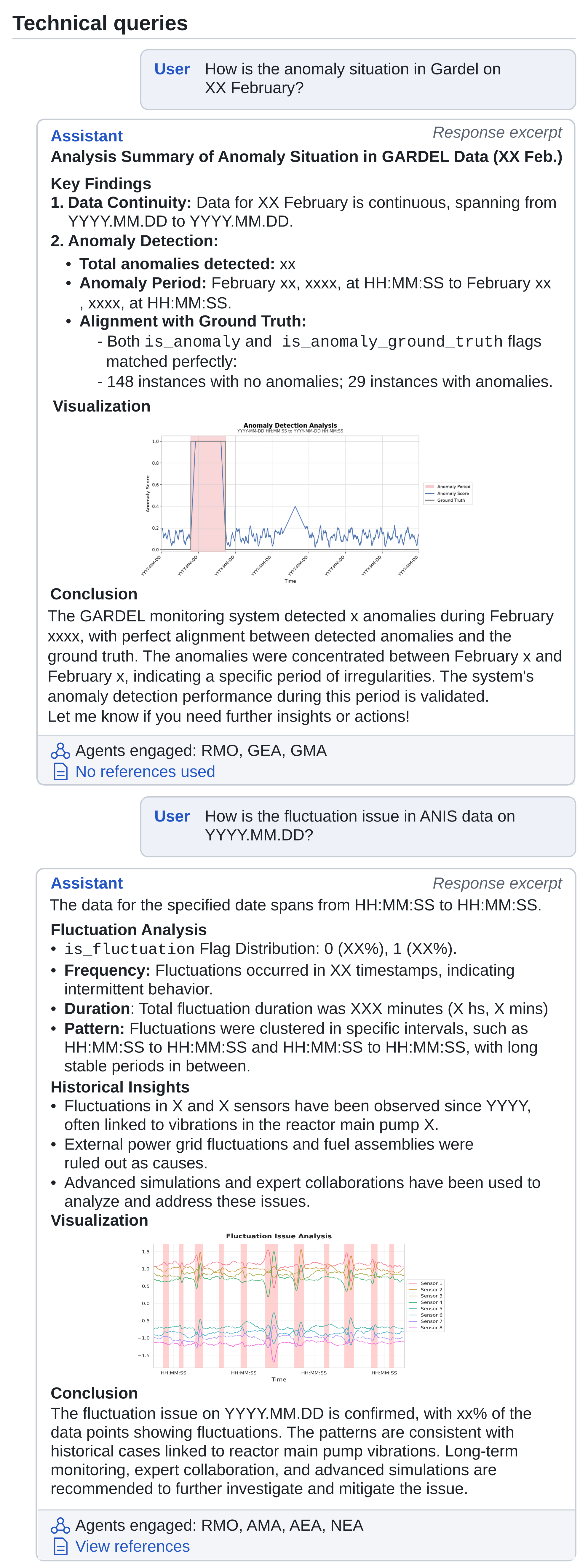}
    \caption{Technical queries concerning anomaly detection in Gardel and fluctuations in ANIS data.}
    \label{fig:technical_queries}
\end{figure}

Figure~\ref{fig:assistant_introduction} shows the response to the question ``Who are you?''. The Reactor Monitoring Assistant introduces its scope, including operational data analysis, monitoring of system variables, and support for anomaly investigation. This exchange is handled by the main assistant without requiring a specialised investigation nor retrieved references. In contrast, the technical queries in Figure~\ref{fig:technical_queries} involve several agents, whose contributions depend on the information required to answer the question.
The GARDEL query requires the retrieval and analysis of monitoring data for a given period. The main assistant engages the GARDEL monitoring agent and the GARDEL expert agent to examine the records for the requested period and interpret them. The response reports the continuity of the data, the number of detected anomalies and the interval in which they occur, and the agreement between the detector output and the ground-truth labels available for that period, where groud-truth labels indicate the ingested anomalies. A plot of the anomaly score against the ground truth accompanies the response, so the engineer can inspect the detected period directly rather than relying only on the summary. The assistant provides advisory support, and plant personnel retain responsibility for operational judgement and action.
The ANIS query exercises a broader investigation involving the main assistant, the ANIS monitoring agent, the ANIS expert agent, and the nuclear expert agent. The question is deliberately vague: it refers to a ``fluctuation issue'' on a given date without specifying the affected sensors, a detection criterion, or a suspected cause. Answering it therefore requires interpreting the request, locating relevant context in domain procedure, and examining the operational data.
Historical reports provide context by describing similar fluctuations and the explanations considered in earlier investigations. The expert agent retrieves this documentation, while the monitoring agent examines the ANIS records for the requested period. The investigation characterises the occurrence, duration, and temporal distribution of fluctuations and produces a plot for inspection. The nuclear expert agent contributes to the interpretation of these observations in relation to the documented plant context. The main assistant combines these contributions into a response linking the observed behaviour to the earlier investigation.
This comparison also supports constructive feedback on the historical report. By checking what occurred during the requested period, the assistant can assess whether the observed pattern is consistent with the documented issue and identify aspects requiring further examination. Similarity with an earlier case provides context for the investigation, establishing a causal explanation requires additional engineering evidence. 

These scenarios illustrate task-dependent agent coordination and the presentation of evidence for engineering assessment. They provide a qualitative demonstration of the workflow, including interpretation of an underspecified request, retrieval across heterogeneous sources, data analysis, visualisation, and synthesis. A systematic assessment of retrieval accuracy, analytical correctness, and usefulness to plant engineers requires evaluation across a broader set of investigations.

\section{Conclusion and Next Steps}
The growing capabilities of LLMs open new possibilities for engineering assistance. Developing these capabilities into a working industrial system requires an architecture designed around the application, its information sources, and the procedures followed by plant personnel.
This paper presented an orchestrated multi-agent assistant for reactor monitoring at the Beznau Nuclear Power Plant. Its development addressed the requirements of a highly regulated operational environment and the complexity of a legacy plant shaped by decades of incremental modernisation. The functional, software, and user-facing architecture was illustrated through a case study. The assistant assembled evidence from several sources, supported interpretation of the event, and proposed follow-up actions with references to the relevant documentation and responsible personnel. These results demonstrate how coordinated agents can support an investigation while keeping the evidence available for engineering review.
The framework is currently deployed in a pilot environment at Beznau, providing an opportunity to assess its usefulness in routine engineering work. Future work will extend the evaluation to a broader range of monitoring scenarios and examine the reliability of agent coordination, the support provided for generated interpretations, and the effort required to maintain domain procedures as plant systems evolve.

\section*{DECLARATION OF GENERATIVE AI AND AI-ASSISTED TECHNOLOGIES IN THE WRITING PROCESS}
During the preparation of this work, the authors used Claude and ChatGPT to improve the manuscript's language and readability. After using these tools, the authors reviewed and edited the content as needed and take full responsibility for the content of the publication

\bibliography{ifacconf}             
                                                   







\appendix
\end{document}